\documentclass[apj]{emulateapj}

\usepackage{epsfig}
\usepackage{amsmath}
\usepackage{enumerate}
\usepackage{natbib}
\usepackage{color}

\newcommand{\nustar}{{\it NuSTAR}}
\newcommand{\ginga}{{\it Ginga}}
\newcommand{\asca}{{\it ASCA}}
\newcommand{\suzaku}{{\it Suzaku}}
\newcommand{\chandra}{{\it Chandra}}
\newcommand{\xmm}{{\it XMM-Newton}}
\newcommand{\swift}{{\it Swift}}
\newcommand{\integral}{{\it INTEGRAL}}
\newcommand{\rxte}{{\it RXTE}}

\newbox\grsign \setbox\grsign=\hbox{$>$} \newdimen\grdimen \grdimen=\ht\grsign
\newbox\simlessbox \newbox\simgreatbox \newbox\simpropbox
\setbox\simgreatbox=\hbox{\raise.5ex\hbox{$>$}\llap
     {\lower.5ex\hbox{$\sim$}}}\ht1=\grdimen\dp1=0pt
\setbox\simlessbox=\hbox{\raise.5ex\hbox{$<$}\llap
     {\lower.5ex\hbox{$\sim$}}}\ht2=\grdimen\dp2=0pt
\setbox\simpropbox=\hbox{\raise.5ex\hbox{$\propto$}\llap
     {\lower.5ex\hbox{$\sim$}}}\ht2=\grdimen\dp2=0pt

\def\simlt{\mathrel{\copy\simlessbox}}

\def\RWD{R_{\rm WD}}
\def\MWD{M_{\rm WD}}

\newcommand{\nuskybgd}{{\ttfamily nuskybgd}}
\newcommand\na{New A}

\shorttitle{NuSTAR and Swift observations of AE~Aqr}
\shortauthors{Kitaguchi et al.}

\begin{document}

\title{NuSTAR and Swift observations of the fast rotating magnetized white dwarf AE~Aquarii}

\author{
Takao~Kitaguchi\altaffilmark{1},
Hongjun~An\altaffilmark{2},
Andrei~M.~Beloborodov\altaffilmark{3},
Eric~V.~Gotthelf\altaffilmark{3},
Takayuki~Hayashi\altaffilmark{4}
Victoria~M.~Kaspi\altaffilmark{2},
Vikram~R.~Rana\altaffilmark{5},
Steven~E.~Boggs\altaffilmark{6},
Finn~E.~Christensen\altaffilmark{7},
William~W.~Craig\altaffilmark{6,8},
Charles~J.~Hailey\altaffilmark{3},
Fiona~A.~Harrison\altaffilmark{5},
Daniel~Stern\altaffilmark{9},
Will~W.~Zhang\altaffilmark{10}
}

\altaffiltext{1}{RIKEN Nishina Center, 2-1 Hirosawa, Wako, Saitama 351-0198, Japan}
\altaffiltext{2}{Department of Physics, McGill University, Rutherford Physics Building, 3600 University Street, Montreal, Quebec, H3A 2T8, Canada}
\altaffiltext{3}{Physics Department and Columbia Astrophysics Laboratory, Columbia University, New York, NY 10027}
\altaffiltext{4}{The Institute of Space and Astronautical Science/JAXA, 3-1-1 Yoshinodai, Chuo-ku, Sagamihara 252-5210, Japan}
\altaffiltext{5}{Cahill Center for Astronomy and Astrophysics, California Institute of Technology, Pasadena, CA 91125}
\altaffiltext{6}{Space Sciences Laboratory, University of California, Berkeley, CA 94720}
\altaffiltext{7}{DTU Space - National Space Institute, Technical University of Denmark, Elektrovej 327, 2800 Lyngby, Denmark}
\altaffiltext{8}{Lawrence Livermore National Laboratory, Livermore, CA 94550}
\altaffiltext{9}{Jet Propulsion Laboratory, California Institute of Technology, Pasadena, CA 91109}
\altaffiltext{10}{NASA Goddard Space Flight Center, Greenbelt, MD 20771}

\begin{abstract}
AE~Aquarii is a cataclysmic variable with the fastest known rotating
magnetized white dwarf ($P_{\rm spin} = 33.08$~s).
Compared to many intermediate polars, AE Aquarii shows a soft X-ray spectrum
with a very low luminosity ($L_{\rm X} \sim 10^{31}$~erg~s$^{-1}$).
We have analyzed overlapping observations of this system with the
\nustar\ and the \swift\ X-ray observatories in September of 2012.
We find the 0.5--30~keV spectra to be well fitted by either
an optically thin thermal plasma model with three temperatures of
$0.75^{+0.18}_{-0.45}$, $2.29^{+0.96}_{-0.82}$, and $9.33^{+6.07}_{-2.18}$~keV,
or
an optically thin thermal plasma model with two temperatures
of $1.00^{+0.34}_{-0.23}$ and $4.64^{+1.58}_{-0.84}$~keV plus a power-law
component with photon index of $2.50^{+0.17}_{-0.23}$.
The pulse profile in the 3--20~keV band is broad and approximately
sinusoidal, with a pulsed fraction of $16.6 \pm 2.3$\%.
We do not find any evidence for a previously reported sharp feature
in the pulse profile.
\end{abstract}


\section{Introduction}

AE Aquarii (hereafter AE~Aqr) is a cataclysmic variable binary system
classified as a member of the DQ Herculis or intermediate polar (IP) class
\citep{1994PASP..106..209P}.
It is a non-eclipsing close binary system at a distance of
$102^{+42}_{-23}$~pc \citep{1997NewA....2..319F},
consisting of a magnetic white dwarf (primary) and a K4--5 V star (secondary)
with an orbital period, $P_{\rm orbit} = 9.88$~hour.
The 33.08~s period makes AE Aqr the fastest-spinning magnetic white dwarf.
The pulsations were originally discovered
in the optical \citep{1979ApJ...234..978P},
then confirmed in soft X-rays \citep{1980ApJ...240L.133P}
and the ultraviolet \citep{1994ApJ...433..313E}.
In the DQ Herculis class, the white dwarf is generally thought to possess
a magnetic field ($B \sim 10^{5-7}$~G) strong enough to channel
the accretion flow from the secondary to the poles of the white dwarf.
Accordingly, hard X-rays are produced by the shock-heated gas, which
reaches temperatures of a few tens of keV near the surface.
The white dwarf photosphere, heated by the hard X-rays, emits ultraviolet
light.
These emissions exhibit spin modulation caused by the varying aspect
of the accreting poles as the white dwarf rotates
(see \citealt{1994PASP..106..209P} for a review).

AE~Aqr stands out as an unusual member of the DQ Herculis class.
It displays strong flares of broad-band emission, from radio to X-rays.
In addition, possible TeV $\gamma$-ray flares have been reported
\citep{1994ApJ...434..292M},
although they have not yet been confirmed with more recent TeV Cerenkov
telescopes \citep{2012MmSAI..83..651M}.
The persistent X-ray luminosity of AE~Aqr ($\sim 10^{31}$~erg~s$^{-1}$)
is two orders of magnitude lower than that of typical IPs.
Its X-ray spectrum has been modeled as emission from an optically thin thermal
plasma with several temperature components, similar to those seen from
other IPs.
However, the highest temperature found in such models was
4.6~keV \citep{2006ApJ...639..397I},
which is significantly lower than the average $kT\approx 22$~keV found in the
22 IPs detected by \integral\ \citep{2009MNRAS.392..630L}.
For these reasons, the mechanism and location of the X-ray emission
are still uncertain
\citep[e.g.,][]{1999ApJ...525..399C,2006ApJ...639..397I,2009ApJ...706..130M}.

Another intriguing feature reported by a \suzaku\ observation
in 2005 is that AE~Aqr may emit non-thermal hard X-rays with
a very narrow pulse profile at the spin period \citep{2008PASJ...60..387T},
suggesting that AE~Aqr may accelerate charged particles in a fashion similar
to rotation-powered pulsars (e.g., \citealt{2006csxs.book..279K} for a review).
However, the \suzaku\ observation in 2006 did not reproduce the earlier
result, leaving the detection of non-thermal X-rays from AE~Aqr uncertain.

In this paper, we present broad-band X-ray observations of AE~Aqr
obtained with \nustar\ and \swift.
Section~\ref{sec:ObsRed} details the observations, data reduction,
and background modeling.
These more sensitive observations in the hard X-ray band can help measure
the maximum temperature of the thermal plasma in AE~Aqr and test the
presence of any beamed nonthermal component.
We describe the spectral modeling in Section~\ref{sec:SpeAna} and
the timing analysis in Section~\ref{sec:TimAna}.
The results and a possible interpretation are discussed in
Section~\ref{sec:Dis}.
Throughout this paper, all errors are given at the 90\% confidence level
unless otherwise stated.

\section{Observations and Data Reductions} \label{sec:ObsRed}

\subsection{NuSTAR}

\subsubsection{Observation Log and Data Screening}

We observed AE~Aqr with \nustar\ \citep{2013ApJ...770..103H},
the first focusing hard X-ray (3--79~keV) telescopes in orbit,
from 2012 September~4, 19:20~UT to September~7, 18:50~UT.
\nustar\ was operated in its default mode throughout the observation.
The acquired data with the observation ID of 30001120 were processed
and screened in the standard way using the \nustar\ pipeline software,
NuSTARDAS version 1.1.1, with
the \nustar\ calibration database (CALDB) version 20130509.
The data were filtered for intervals of high background, including
Earth occultations and South Atlantic Anomaly (SAA) passages.
This resulted in a total of 125~ks of dead-time corrected exposure time.

All photon arrival times were corrected to the Solar System barycenter
using the JPL DE200 ephemeris and
the \chandra-derived coordinates
(20:40:09.185,~$-00$:52:15.08; J2000) which have sub-arcsecond uncertainties.
The source photons were extracted from a circular region of
radius $1\farcm0$ centered on AE~Aqr.
The source spectra were grouped with a minimum of 50~counts per bin.
The background spectra were generated with the background modeling
tool, \nuskybgd\ (Wik et al. in preparation).
A detailed description of the background modeling is given in
Section~\ref{sec:bgd}.
The telescope response files, ARF and RMF were also generated by
NuSTARDAS.

\subsubsection{Background Spectral Modeling} \label{sec:bgd}

Since AE~Aqr is a faint source in hard X-rays, the \nustar\ background
spectrum must be carefully subtracted from the source spectrum.
Blank-sky observations show the background rate varies in an energy
dependent way with detector position by a factor of $\sim2$
\citep{2013ApJ...770..103H}.
In addition, the internal background rate varies in time due to changes
in the cosmic radiation intensity associated with the geomagnetic cut-off
rigidity, as well as the elapsed time since SAA passage.

In order to perform accurate background subtraction, the background
spectrum was empirically modeled using the \nustar\ background fitting
and modeling tool, \nuskybgd\
(revision 52).
\nuskybgd\ fits blank-sky spectra in user-selected regions
of the same observation,
and then generates a background spectrum within any regions
with the best-fit parameters (Wik et al. in preparation for a more detailed).

\begin{figure}
 \centering
 \includegraphics[angle=0,scale=0.43]{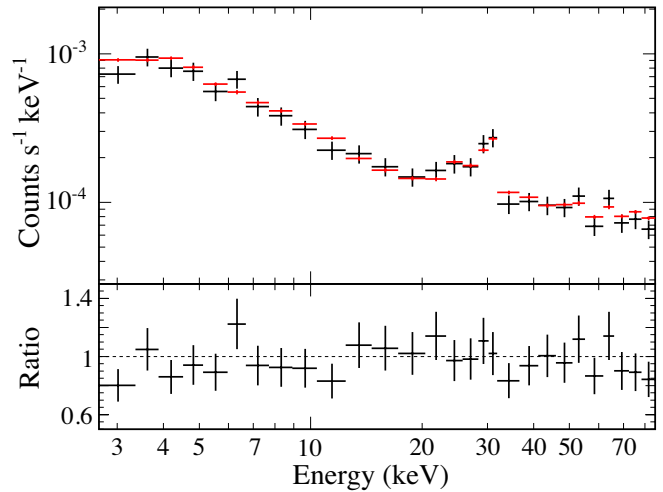}
 \caption{Comparison of (red) the background model produced by \nuskybgd\ with
 (black) the actual blank-sky spectrum in the AE~Aqr observation.
 The bottom panel shows the spectral ratio of the actual data to the model.
 \label{fig:CompBGD}}
\end{figure}

Three blank-sky spectra for each telescope were extracted from
annular regions of radius $120''-270''$, $270''-370''$, and
$370''-740''$ centered on AE~Aqr to model the background spectra of
the two \nustar\ telescopes.
The blank-sky spectra were well fitted by the model with
$\chi^{2}/{\rm dof} = 1217.6/1124$.
The background count spectrum in the source region was modeled with
an exposure 20 times longer than the actual one to reduce statistical
errors.
In order to verify the background model, the background spectrum
in a blank-sky region $3.5'$ from the source in a northeasterly
direction was simulated and compared to the actual spectrum extracted
from the same region.
The ratio of the actual spectrum to the model in Figure~\ref{fig:CompBGD}
was fitted by a constant factor of $0.98 \pm 0.04$ with
$\chi^{2}/{\rm dof} = 15.7/27$, showing the background model is consistent
with the actual spectrum within statistical errors.

\begin{figure}
 \centering
 \includegraphics[angle=0,scale=0.44]{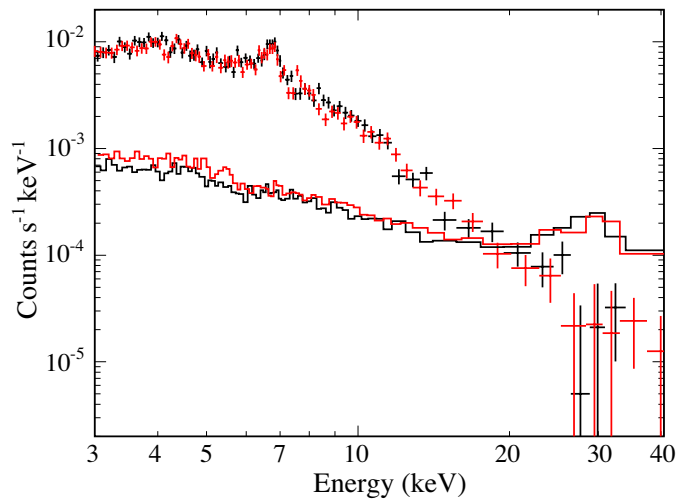}
 \caption{
 X-ray spectra of AE~Aqr observed with the two \nustar\ telescopes,
 (black) FPMA and (red) FPMB.
 The cross points and solid histograms are the background-subtracted
 spectra and background models, respectively.
 \label{fig:spec}}
\end{figure}

Figure~\ref{fig:spec} shows the AE~Aqr spectra obtained with
the two \nustar\ hard X-ray telescopes (FPMA and FPMB)
and the background models.
The highly ionized iron line can be seen around 6.7~keV.
The total count rate in the 3--30~keV energy band after the background
subtraction and dead-time correction is
$(9.2 \pm 0.1) \times 10^{-2}$~counts~s$^{-1}$.

\subsection{Swift}

\swift\ \citep{2004ApJ...611.1005G} observed AE~Aqr simultaneously with
\nustar\ from 2012 September~6, 06:07 to 07:53 UT.
The \swift\ X-ray telescope \citep[XRT;][]{2005SSRv..120..165B}
was operated in the photon counting mode during the observation.
The XRT data (observation ID 00030295037) were processed in the
standard manner with xrtpipeline (ver. 0.12.6) in HEASoft (ver. 6.13).
The total exposure time after the data screening is 1.54~ks.
The source photons were extracted from a circular region of radius
$1\farcm2$ centered on AE~Aqr.
The source spectrum was rebinned to have at least 20~counts per bin.
The background spectrum was extracted from an annular region of radius
$1\farcm7 - 7\farcm0$ centered on the source and was scaled by an area
ratio of the source region to the background region.

\section{Spectral Analysis} \label{sec:SpeAna}

\subsection{Spectral Fitting with Multi-Temperature Models} \label{sec:mulTfit}

The X-ray spectra of AE~Aqr observed with \asca\
\citep{1999ApJ...525..399C}, \xmm\ \citep{2006ApJ...639..397I},
and \suzaku\ \citep{2008PASJ...60..387T}
have been modeled using an optically thin thermal plasma emission
model with a few different temperature components in the same way
as for other IPs.
Therefore in the joint fitting of \swift\ and \nustar\ spectra, we
adopted an emission model from a collisionally ionized diffuse plasma,
{\ttfamily APEC} \citep{2001ApJ...556L..91S}.
Each {\ttfamily APEC} model was constrained to have common metal abundances
relative to solar from \cite{2000ApJ...542..914W}, but was allowed
to have different temperatures and normalizations.
We use the {\ttfamily tbabs} model \citep{2000ApJ...542..914W} to account for
interstellar and self-absorption.
In addition, the cross-normalization factors of XRT/FPMA and FPMB/FPMA
were allowed to vary.

\begin{figure}
 \centering
 \includegraphics[angle=0,scale=0.33]{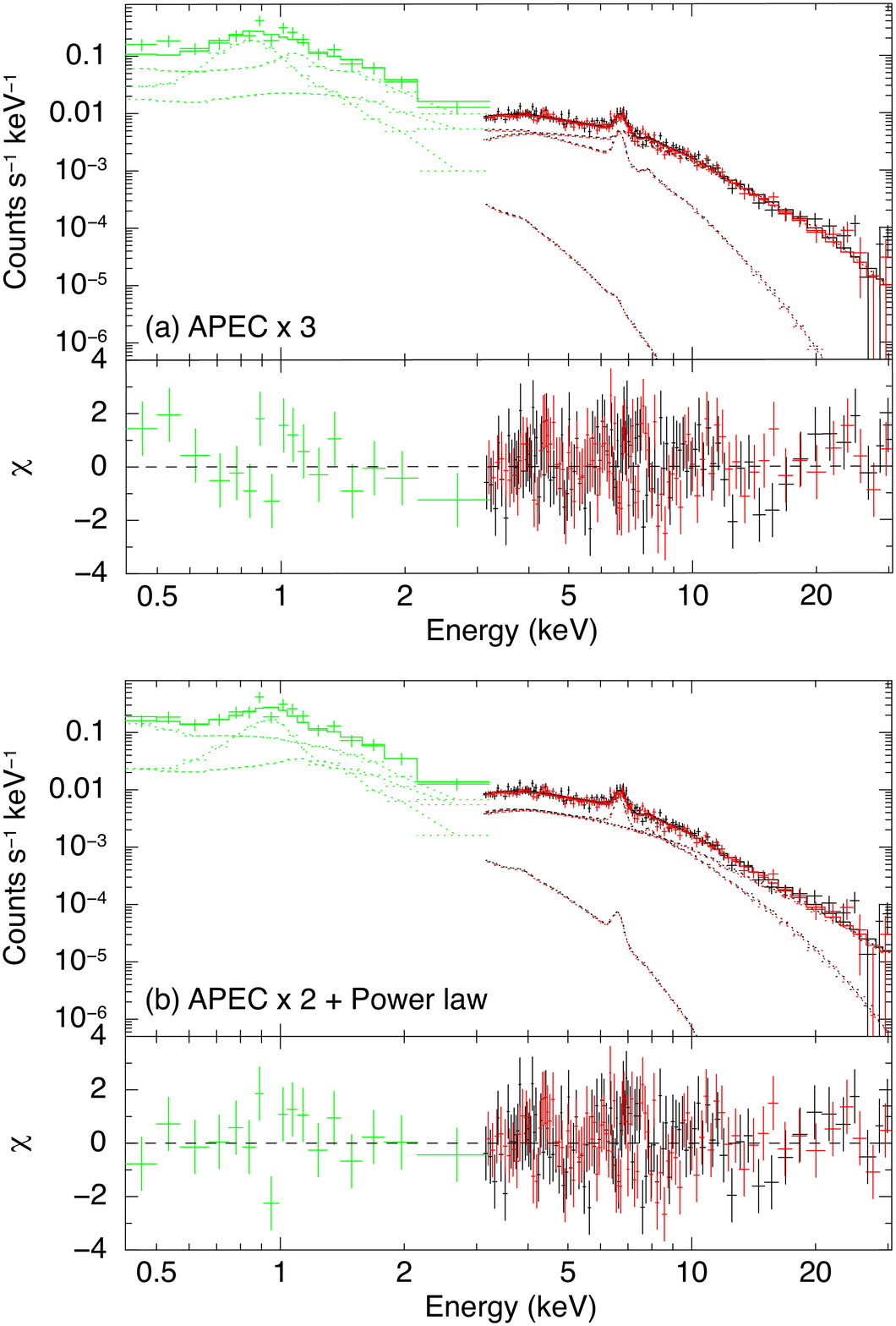}
 \caption{X-ray Spectra of (green) \swift/XRT, (black) \nustar/FPMA,
 and (red) \nustar/FPMB overlaid with (a) the best-fit
 three-temperature {\ttfamily APEC} model and (b) two-temperature model with a
 power-low emission.
 The model components are drawn with dotted lines.
 The bottom panels show fit residuals in terms of $\sigma$ with
 error bars of size one.
 \label{fig:Spec3ApecFit}}
\end{figure}

In order to determine the number of {\ttfamily APEC} components with
different temperatures, we added new {\ttfamily APEC} components one by one
until the fit was not significantly improved, as determined by the F-test.
Improvement of the fit is significant with the chance probability of
$\ll1$\% until the number of components is increased to three,
whereas little improvement was found by the addition of a fourth
component with the chance probability of 8.4\% derived from the $\chi^{2}$
reduction from 247.1 (dof = 201) to 241.0 (dof = 199).
The 0.5--30~keV spectra of \swift\ and \nustar\ with the best-fit
three-temperature {\ttfamily APEC} models are shown in Figure~\ref{fig:Spec3ApecFit}~(a),
the parameters of which are listed in Table~\ref{tbl:Spec3ApecFit}.
with previous results for reference.
The highest temperature, $9.3^{+6.1}_{-2.2}$~keV, is considerably
higher than previously observed for this source (4.6~keV from
\citealt{2006ApJ...639..397I}),
and approaches the average temperature of 22~keV for any other IPs
\citep{2009MNRAS.392..630L}.
The metal abundance, $0.76^{+0.17}_{-0.13}~Z_{\odot}$, is consistent
with that determined by \xmm\ and \suzaku.
The total luminosity in 0.5--10~keV is
$9.3^{+1.0}_{-2.2} \times 10^{30}$~erg~s$^{-1}$ for an assumed distance
of 100~pc.

\begin{deluxetable}{lcc}
 \tablewidth{0pt}
 \tablecaption{Best-fit parameters of the three-temperature model.
 \label{tbl:Spec3ApecFit}}
 \tablewidth{0pt}
 \tablehead{
 \colhead{Parameter} & \colhead{Unit} & \colhead{Value}
 }
 \startdata
 $N_{\rm H}$ & ($10^{20}$~cm$^{-2}$)   & $<82.2$ \\
 $Z$        & ($Z_{\odot}$\tablenotemark{a}) & $0.76^{+0.17}_{-0.13}$  \\
 $kT_{1}$    & (keV)                    & $0.75^{+0.18}_{-0.45}$  \\
 $F_{1}$\tablenotemark{b} & ($10^{-12}$~erg~s$^{-1}$~cm$^{-2}$) & $1.97^{+0.94}_{-0.85}$  \\
 $kT_{2}$    & (keV)                    & $2.29^{+0.96}_{-0.82}$  \\
 $F_{2}$\tablenotemark{b} & ($10^{-12}$~erg~s$^{-1}$~cm$^{-2}$) & $3.53^{+0.55}_{-0.64}$  \\
 $kT_{3}$    & (keV)                    & $9.33^{+6.07}_{-2.18}$  \\
 $F_{3}$\tablenotemark{b} & ($10^{-12}$~erg~s$^{-1}$~cm$^{-2}$) & $2.26^{+0.99}_{-0.96}$  \\
 $C_{\rm FPMB}/C_{\rm FPMA}$\tablenotemark{c} &          & $1.00 \pm 0.03$ \\
 $C_{\rm XRT}/C_{\rm FPMA}$\tablenotemark{d}  &          & $0.94^{+0.19}_{-0.22}$ \\
 $\chi^{2}/{\rm dof}$ &                 & 247.1/201
 \enddata
 \tablenotetext{a}{Solar abundance from \cite{2000ApJ...542..914W}, in which
 the solar iron abundance relative to hydrogen is $2.69 \times 10^{-5}$.
 Note that the solar abundance from \cite{1989GeCoA..53..197A}, in which
 the solar iron abundance is $4.68 \times 10^{-5}$, is employed in other
 AE~Aqr articles
 (e.g. \citealt{2006ApJ...639..397I,2008PASJ...60..387T,2012MNRAS.421.1557O}).}
 \tablenotetext{b}{Unabsorbed flux in 0.5--10~keV.}
 \tablenotetext{c}{Cross-normalization factor between \nustar/FPMB and \nustar/FPMA.}
 \tablenotetext{d}{Cross-normalization factor between \swift/XRT and \nustar/FPMA.}
\end{deluxetable}

Although other IP spectra commonly contain a strong neutral
(or low-ionized) iron line around 6.4~keV with a comparable
equivalent width to highly ionized iron emission around
6.7~keV \citep[e.g.,][]{1999ApJS..120..277E,2010A&A...520A..25Y},
we do not detect this feature clearly in the AE~Aqr spectrum.
In fact, \xmm\ spectra constrained the upper limit to the 6.4~keV
equivalent width to be $<61$~eV \citep{2006ApJ...639..397I}.
We tried to add a narrow Gaussian emission model with a center energy
fixed at 6.4~keV to the three-temperature {\ttfamily APEC} model, and simultaneously
fit the spectra of \swift\ and \nustar\ with it.
The intensity of the neutral iron line is
$1.33_{-1.33}^{+1.26} \times 10^{-14}$~erg~s$^{-1}$~cm$^{-2}$,
which corresponds to an equivalent width of $37.9_{-37.9}^{+41.7}$~eV.
On the other hand, the intensity of the highly ionized iron line around
6.7~keV was determined to be
$1.11_{-0.09}^{+0.11} \times 10^{-13}$~erg~s$^{-1}$~cm$^{-2}$,
or an equivalent width of $495_{-53}^{+365}$~eV, with the line central
energy of $6.732_{-0.014}^{+0.063}$~keV.
Therefore, the neutral iron line is much weaker than the highly
ionized iron line.

\subsection{Spectral Fitting with Non-Thermal Emission} \label{sec:NT}

In order to search for possible non-thermal hard X-rays as suggested
by \cite{2008PASJ...60..387T}, we simultaneously fitted the spectra
of \swift\ and \nustar\ with a multi-temperature plus power-law
emission model.
In the same way as the determination of the number of {\ttfamily APEC} components
described above, we first fit the spectra using a single {\ttfamily APEC} with
a power-law model,
and then added new {\ttfamily APEC} models one by one until the fit was not
significantly improved as determined by the F-test.
We find two-temperatures with power-law emission is the statistically
favored model.

The \swift\ and \nustar\ spectra with the best-fit model are shown
in Figure~\ref{fig:Spec3ApecFit}~(b), the parameters of which are listed
in Table~\ref{tbl:Spec2ApecPLFit}.
Compared with the three-temperature {\ttfamily APEC} model, the fit with the
two-{\ttfamily APEC} model with the power-law emission is slightly but not
significantly preferred.
We cannot distinguish whether the hard X-ray component detected with
\nustar\ is thermal or non-thermal emission.

\begin{deluxetable}{lcc}
 \tablewidth{0pt}
 \tablecaption{Best-fit parameters of the two-temperature model with
 power-law emission.
 \label{tbl:Spec2ApecPLFit}}
 \tablewidth{0pt}
 \tablehead{
 \colhead{Parameter} & \colhead{Unit} & \colhead{Value}
 }
 \startdata
 $N_{\rm H}$ & ($10^{20}$~cm$^{-2}$)   & $<82.2$ \\
 $Z$         & ($Z_{\odot}$) & $1.14^{+0.62}_{-0.33}$ \\
 $kT_{1}$    & (keV)                    & $1.00^{+0.34}_{-0.23}$ \\
 $F_{1}$\tablenotemark{a} & ($10^{-12}$~erg~s$^{-1}$~cm$^{-2}$) & $2.01^{+1.25}_{-1.03}$  \\
 $kT_{2}$    & (keV)                    & $4.64^{+1.58}_{-0.84}$ \\
 $F_{2}$\tablenotemark{a} & ($10^{-12}$~erg~s$^{-1}$~cm$^{-2}$) & $1.86^{+0.96}_{-0.81}$  \\
 $\Gamma$    &                          & $2.50^{+0.17}_{-0.23}$ \\
 $F_{\rm PL}$\tablenotemark{b} & ($10^{-12}$~erg~s$^{-1}$~cm$^{-2}$) & $4.07^{+1.66}_{-1.50}$ \\
 $C_{\rm FPMB}/C_{\rm FPMA}$ &          & $1.00 \pm 0.03$ \\
 $C_{\rm XRT}/C_{\rm FPMA}$  &          & $0.81^{+0.22}_{-0.19}$ \\
 $\chi^{2}/{\rm dof}$ &                 & 244.5/201
 \enddata
 \tablenotetext{a}{Unabsorbed flux of each thermal emission component in 0.5--10 keV}
 \tablenotetext{b}{Unabsorbed flux of the power-law model in 0.5--10 keV}
\end{deluxetable}

\section{Timing Analysis} \label{sec:TimAna}

\subsection{Spin Period Determination}

We examined the hard X-ray pulse profile for evidence of the narrow
pulsation reported by \suzaku\ \citep{2008PASJ...60..387T}.
Previous measurements have shown that soft X-rays below 10~keV
are sinusoidally modulated with a single peak in the spin phase
of the white dwarf (e.g., \citealt{2006MNRAS.369.1983M}).
In this case, the $Z^{2}_{1}$-statistic, or Rayleigh test
\citep{1983A&A...128..245B} is more sensitive to determine
the spin period than the epoch folding technique with the $\chi^{2}$
test \citep{1983ApJ...272..256L}.

\begin{figure}
 \centering
 \includegraphics[angle=0,scale=0.42]{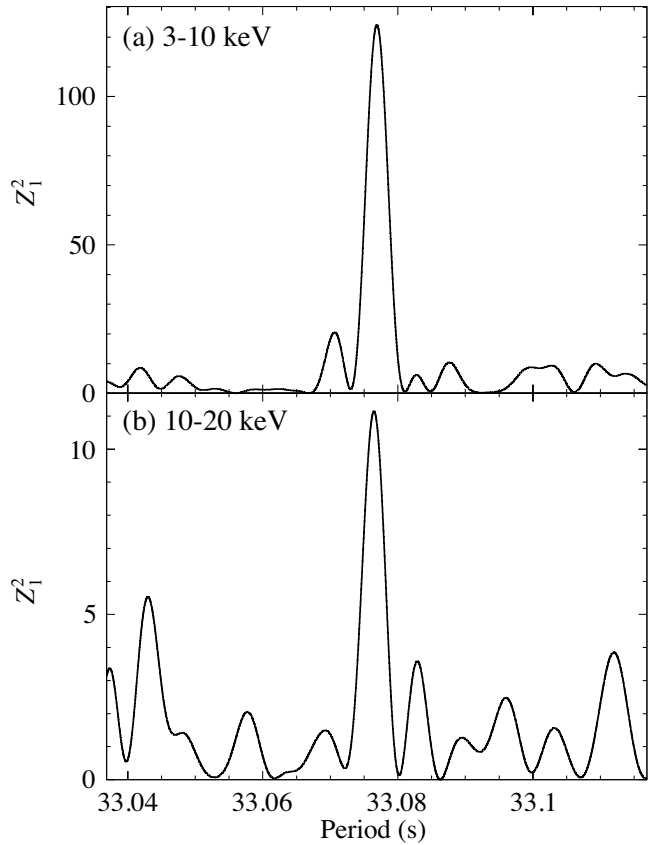}
 \caption{$Z^{2}_{1}$ periodograms acquired with \nustar\ in the energy
 bands of (a) 3--10 and (b) 10--20~keV.
 \label{fig:period}}
\end{figure}

Figure~\ref{fig:period}~(a) shows the $Z^{2}_{1}$ periodogram
using \nustar\ 3--10~keV data with barycenter-corrected time.
We found the peak at $33.0769 \pm 0.0004$~s with $Z^{2}_{1} = 124.1$.
Within the uncertainty, the period is consistent with previously
measured values in the optical band
\citep{1979ApJ...234..978P,1994MNRAS.267..577D},
and at X-ray energies
\citep{1999ApJ...525..399C,2006ApJ...646.1149C,2006MNRAS.369.1983M,2008PASJ...60..387T,2012MNRAS.421.1557O}.

We also searched for a signal in the 10--20~keV energy band.
The hard X-rays above 20~keV were ignored to improve the signal-to-noise
ratio since the source rate from AE~Aqr falls below that of the background
in that energy range (Figure~\ref{fig:spec}).
Evidence for a pulsation was found also in this band at the expected pulse
period, $33.0765 \pm 0.0016$~s, with  $Z^{2}_{1} = 11.1$,
corresponding to a null hypothesis probability of 0.38\%,
or $2.9\sigma$ detection significance [Figure~\ref{fig:period}~(b)].

\subsection{Pulse Profile}

\begin{figure}
 \centering
 \includegraphics[angle=0,scale=0.40]{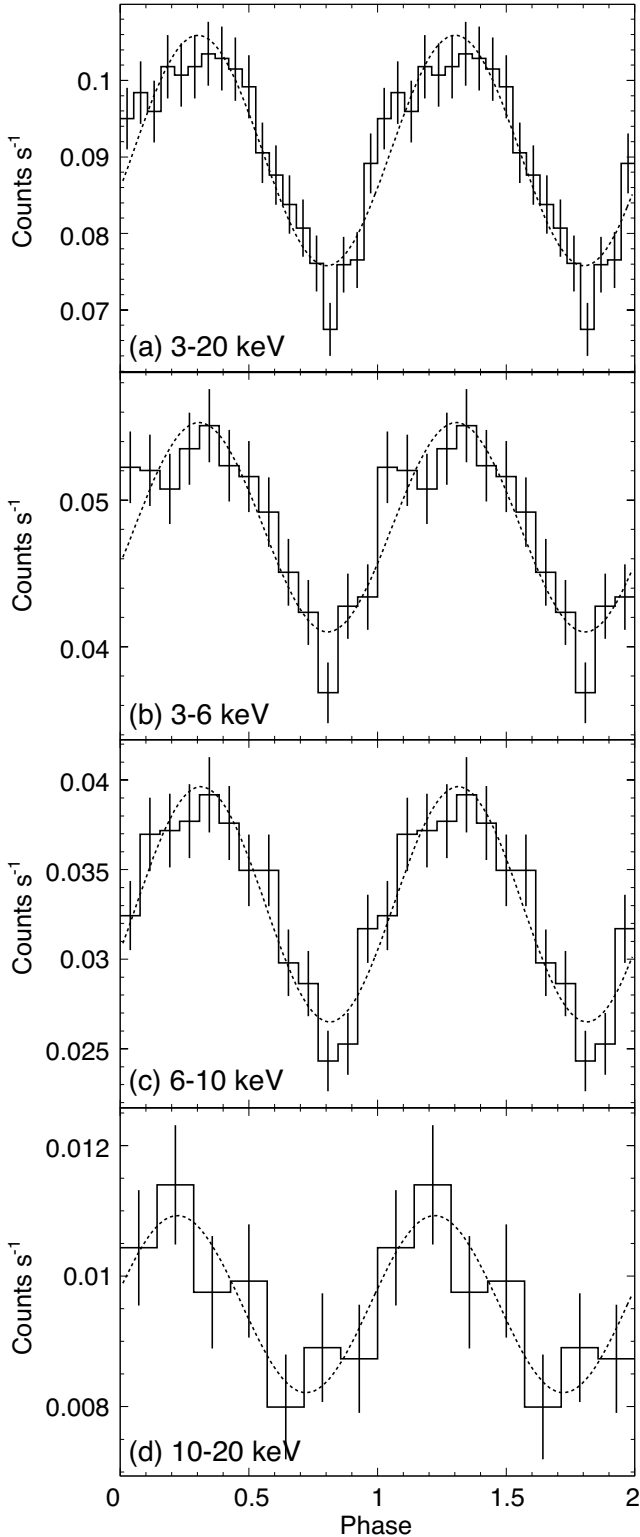}
 \caption{AE~Aqr pulse profiles folded on the best period of 33.0769~s
 in (a) 3--20, (b) 3--6, (c) 6--10, and (d) 10--20~keV energy bands.
 The dashed curves show the best-fit sinusoidal models.
 The epoch of phase 0 corresponds to 56174.5 in MJD.
 \label{fig:prof}}
\end{figure}

Figure~\ref{fig:prof} presents the pulse profiles in several energy
bands, folded on the best determined period,
with the background subtracted using the model of Section~\ref{sec:bgd}.
The profiles are well represented by sinusoidal functions
(see Table~\ref{tbl:prof}).
We find no evidence for significant variation in the phase or
relative modulation amplitude (or pulse fraction) with energy.
The mean pulse fraction, $16.6\pm2.3\%$, is consistent with that found
in the quiescent phase from \xmm\ observations
\citep{2006ApJ...639..397I,2006ApJ...646.1149C}.

\begin{deluxetable}{lccc}
 \tablewidth{0pt}
 \tablecaption{Pulse Profile Parameters\tablenotemark{a} for AE~Aqr.
 \label{tbl:prof}}
 \tablehead{
 \colhead{Energy range} &
 \colhead{Peak phase\tablenotemark{b}} &
 \colhead{Pulse fraction} &
 \colhead{$\chi^{2}/{\rm dof}$} \\
 \colhead{(keV)} &
 \colhead{} &
 \colhead{(\%)} &
 \colhead{}
 }
 \startdata
 3--20   & $0.30 \pm 0.02$ & $16.6 \pm 2.3$ & 17.2/16 \\
 \tableline
 3--6    & $0.31 \pm 0.04$ & $14.8 \pm 3.1$ & 11.7/10 \\
 6--10   & $0.31 \pm 0.03$ & $19.9 \pm 3.7$ & 8.4/10 \\
 10--20  & $0.22 \pm 0.10$ & $14.2 \pm 8.0$ & 2.5/4
 \enddata
 \tablenotetext{a}{Best-fit parameters for a fit to the \nustar\ pulse
 profiles presented in Figure~\ref{fig:prof} using a sinusoid model
 with a constant offset, $C$:
 $C(1.0+f_{\rm P}\cos(2\pi(\phi-\phi_0)))$,
 where $\phi_0$ is the peak phase and $f_{\rm P}$ is the pulse fraction.}
 \tablenotetext{b}{Phase 0 corresponds to that of Figure~\ref{fig:prof}.}
\end{deluxetable}

We searched for a narrow peak in the pulse profile as reported by
\cite{2008PASJ...60..387T} based on the \suzaku\ observation of 2005.
We investigated the pulse profile using different numbers of bins per cycle
ranging from 5 to 50.
Furthermore, we examined the profile using 29 bins per cycle as per
the \suzaku\ analysis and reproduced 1000 pulse profiles, each
having a different start time at phase zero.
Then we looked for any significant outliers from the sinusoidal model
with a 95\% confidence level.
We conclude that there is no evidence for an additional sharp pulse
component in our data.

\section{Discussion} \label{sec:Dis}

\subsection{Comparison of Non-Thermal Hard X-ray Emission with Previous \suzaku\ Results}

The 10--30~keV flux determined with \nustar\ is
$5.0^{+1.3}_{-1.4}~\times~10^{-13}$~erg~s$^{-1}$~cm$^{-2}$, about half
of the best-fit \suzaku/HXD-PIN flux in 2005 \citep{2008PASJ...60..387T}.
The derived power-law index, $2.5 \pm 0.2$, is also inconsistent
with the \suzaku\ value, $1.1 \pm 0.6$,
and is steeper than those found for rotation-powered pulsars,
which range from 0.6 to 2.1 \citep{2003ApJ...591..361G, 2006csxs.book..279K}.
Here we note that the previously reported \suzaku\ fitting errors
did not include the systematic uncertainty
of the \suzaku/HXD-PIN background model because the authors described
that the hard X-ray flux was consistent with the narrow pulse flux
in the spin profile.
However, the systematic uncertainty, 3\% \citep{2009PASJ...61S..17F},
is comparable to the background-subtracted count rate, and therefore
should not be ignored (see Figures~6 and 7 of \citealt{2008PASJ...60..387T}
for comparison of the rates).
It is possible that the discrepancy of the non-thermal emission
parameters is due to the neglected systematic error in
\cite{2008PASJ...60..387T}.

Furthermore, we find no evidence of the narrow pulse profile with
a pulse fraction of nearly 100\% and a duty cycle of $\sim 0.1$
reported by \cite{2008PASJ...60..387T}.
In contrast with the \suzaku\ result, we marginally detect
the sinusoidal modulation in the 10--20~keV energy band at
a significance level of $2.9 \sigma$.
In order to determine whether or not \nustar\ is sensitive to
the narrow pulse profile suggested by \suzaku, we estimated
its count rate considering the \nustar\ telescope response
and assuming the \suzaku\ result.
The expected 10--20~keV count rate within a duty cycle of 0.1 is
$0.086 \pm 0.003$~count~s$^{-1}$, 10 times higher than the
time-averaged rate measured with \nustar,
indicating that \nustar\ would have been capable of significantly
detecting the narrow pulsation.

A possible explanation of the difference between the \suzaku\
and \nustar\ results is that AE~Aqr
may have varied and its hard X-ray flux decreased.
However, the soft X-ray flux determined with \nustar\
in 2012 is consistent with previous observations
made with \suzaku\ in 2005
(see Table~\ref{tbl:Spec3ApecFit}),
within the cross-normalization factor of \nustar\ to \suzaku/XIS
of $\sim 15$\% (based on simultaneous observations of calibration
targets).
It is difficult to explain a mechanism only in hard X-rays that would
make the flux vary without correlation of thermal soft X-rays, a part
of which also modulates with the spin period.

\subsection{Possible Interpretation of the Observed Emission}

Two energy sources could, in principle, power the observed X-ray
luminosity $L_X\sim 10^{31}$~erg~s$^{-1}$: liberation of gravitational
energy of accreting matter and the rotational energy of the white dwarf.

The white dwarf was reported to spin down with a rate
$\dot{P}=5.64\times 10^{-14}$~s~s$^{-1}$ \citep{1994MNRAS.267..577D},
which corresponds to spindown power
$L_{\rm sd}\sim 10^{34}$~erg~s$^{-1}\gg L_X$.
This suggests the possibility that the X-ray emission of the white dwarf
is fed by rotation \citep{2012ARep...56..595I}.
The rotation-induced electric field could accelerate particles to high Lorentz factors
if the magnetosphere is plasma-starved and the electric field is not screened.
Synchrotron emission
was suggested as a possible radiative mechanism in such a model
\citep{2008PASJ...60..387T}.
It is, however, unclear how the observed luminosity, spectrum, and pulse
profile observed by \nustar\ would be produced by relativistic particles.
The efficiency of synchrotron emission is expected to be small,
since the particles are accelerated along the magnetic field lines.
Emission from accelerated particles is also expected to be strongly
beamed along the field lines, which is inconsistent with the observed
broad pulse profile in Figure~\ref{fig:prof}.

Emission powered by accretion is a plausible mechanism.
The standard model of accreting IPs involves an accretion column
heated by the accretion shock and cooled by thermal bremsstrahlung
\citep{1973PThPh..49.1184A}.
It explains well the X-ray spectra of many IPs (see
\citealt{1999MNRAS.306..684C} for \ginga\ data,
\citealt{2005A&A...435..191S} for \rxte,
\citealt{2009MNRAS.392..630L} for \swift/XRT and \integral/IBIS,
\citealt{2009A&A...496..121B} for \swift/BAT,
and \citealt{2010A&A...520A..25Y} for \suzaku).
The main parameters of the model are the
free-fall velocity $v_{\rm ff}=(2G\MWD/\RWD)^{1/2}$
and the accretion rate per unit area,
$a=\dot{M}/4\pi \RWD^2 f$, where $f$ is the fraction of the white dwarf
surface occupied by the accretion column.
The standard model assumes that the shock is radiative and pinned
to the white dwarf surface, $h\ll \RWD$. The shock altitude
$h\simlt vt_{\rm br}$, where $v=v_{\rm ff}/4\sim 10^8$~cm~s$^{-1}$
is the postshock velocity of the accreting gas and
$t_{\rm br}\sim 0.3/a$~s
is the bremsstrahlung cooling timescale (with $a$ expressed in units of
g~s$^{-1}$~cm$^{-2}$).
The net accretion rate $\dot{M}$
is determined from the X-ray luminosity $L_X=G\MWD\dot{M}/\RWD$.

This scenario would imply
$\dot{M}\sim 2\times 10^{14}$~g~s$^{-1}$ for AE~Aqr, and the condition
$a\gg 10^{-2}$~g~s$^{-1}$~cm$^{-2}$ is satisfied if $f \ll 10^{-3}$.
Such a small $f$ may be possible if, e.g., the accretion flow in AE~Aqr
is concentrated in a thin wall of the accretion column
(as \citealt{1976MNRAS.175..395B} suggested for accreting neutron stars).
However, on the assumption of $M_{\rm WD} \sim 0.7$~$M_{\odot}$
determined with the optical measurements
(e.g. \citealt{2006MNRAS.368..637W,2008MNRAS.387.1563E}),
our calculations of the spectrum predicted by the standard accretion model
do not provide a good fit for the \nustar\ data ---
the observed spectrum with the highest temperature of
$9.3^{+6.1}_{-2.2}$~keV is softer than the predicted
postshock temperature of 29 keV.

Two modifications might make the accretion model consistent with
\nustar\ observations.
(1) The shock altitude $h$ in AE~Aqr could be significant,
perhaps even comparable to the white dwarf radius.
Such tall accretion columns with low accretion rates were recently
studied by \cite{2013arXiv1307.7881H}.
The shallower gravitational potential at the high altitude
results in a reduction of the postshock
temperature below 20~keV typical of other IPs.
The tall accretion column in AE~Aqr
was previously proposed as the source of the large hot spots on the white
dwarf surface inferred from observations by the {\it Hubble} telescope
\citep{1994ApJ...433..313E}
(2) The X-ray spectrum emitted by the accretion column
could be affected by additional radiative losses due to cyclotron emission
in the infrared band. Cyclotron emission
is important when $B>10^{6}$~G, and may carry away a significant part
of the accretion column energy, especially at large altitudes where the plasma
has the highest temperature. This effect could explain the relatively soft
extended spectrum of AE~Aqr above 3~keV.
Detailed calculations of such models will be presented elsewhere.

In addition to the low X-ray luminosity and soft spectrum, AE~Aqr has
another special feature
--- weak absorption of soft X-rays, which corresponds to
$N_H< 8\times 10^{21}$~cm$^{-2}$, lower than the $N_H>10^{22}$~cm$^{-2}$
observed in many IPs \citep[e.g.][]{1999ApJS..120..277E,2010A&A...520A..25Y}.
This feature could be explained by the lower density of the accretion column,
and needs to be further investigated with detailed models.
Also note that AE~Aqr does not show evidence for a neutral
(or weakly ionized) iron line.
Among other IPs previously studied in the hard X-ray band, EX~Hydrae
has the lowest luminosity $L_X\sim 10^{32}$~erg~s$^{-1}$ and also
the lowest column with 
$N_H\simlt 10^{22}$~cm$^{-2}$ \citep{2005A&A...435..191S}.
The shape of its 3-30~keV spectrum is not far from that of AE~Aqr,
although EX~Hydrae is a nearly synchronous system
with an orbital period of 98.3~min and a long spin period of 67.0~min
\citep{2005AJ....129.1985B}, unlike AE~Aqr.

The much smaller accretion rate of AE~Aqr compared to many
other IPs is thought to be due to the magnetic propeller effect
\citep{1997MNRAS.286..436W}, whereby most of the mass lost by
the secondary is being flung out of the binary by the magnetic
field of the rapidly rotating white dwarf.
\cite{2004ApJ...614..349N,2008ApJ...672..524N} performed numerical
simulations of accretion flows in magnetic cataclysmic variables
with wide ranges of magnetic moment, $\mu = 10^{32-36}$~G~cm$^{3}$, and
$P_{\rm spin}/P_{\rm orbit} = 10^{-3}-1$.
They demonstrated that IPs with a very small ratio of
$P_{\rm spin}/P_{\rm orbit} < 10^{-2}$ like AE~Aqr
($P_{\rm spin}/P_{\rm orbit} = 9.3 \times 10^{-4}$) show magnetic
propeller effects.
Therefore, the tall accretion column could be formed in IPs with
$P_{\rm spin} / P_{\rm orbit} < 10^{-2}$, though such systems would
also be expected to have low X-ray luminosities, thus making them more
challenging to identify. AE~Aqr is the only currently known example.
It is also possible that nearly synchronous systems similar
to EX~Hydrae could possess a tall accretion column.
The low accretion rate in such
systems is consistent with the picture of ring-like accretion
discussed by \cite{2008ApJ...672..524N}.

\section{Summary}

We have analyzed X-ray data of the magnetized white dwarf AE~Aqr
observed with \nustar\ for 125~ks and \swift\ for 1.5~ks.
The 0.5--30 keV spectra are well characterized by either
an optically thin thermal plasma model with three temperatures
with the highest temperature of $9.33^{+6.07}_{-2.18}$~keV,
or
an optically thin thermal plasma model with two temperatures
plus a power-law component with photon index of $2.50^{+0.17}_{-0.23}$.
The 3--20~keV pulse profile shows a sinusoidal modulation,
with a pulsed fraction of $16.6 \pm 2.3$\%.
We find no evidence for the previously reported sharp
feature in the pulse profile.

The observed soft X-ray spectrum with the broad pulse profile
is hard to explain as a result of rotation-powered emission.
Instead accretion-powered emission is more likely, although
the observed spectrum is softer than predicted by the standard
accretion column model.
We suggest two modifications of the standard model to explain
the AE~Aqr spectrum: the shock altitude could be high, comparable
to the white dwarf radius, and cyclotron emission with $B>10^{6}$~G
could additionally cool down the accretion plasma.
Detailed calculations of such models will hopefully reproduce
the spectrum and pulse profile of AE~Aqr with the white dwarf mass
consistent with that determined with the optical measurements.

\acknowledgments

This work was supported under NASA Contract No. NNG08FD60C, and made
use of data from the \nustar\ mission, a project led by the California
Institute of Technology, managed by the Jet Propulsion Laboratory, and
funded by the National Aeronautics and Space Administration.
We thank the \nustar\ Operations, Software and Calibration teams
and the \swift\ Operations team for
support with the execution and analysis of these observations.
This research has made use of the \nustar\ Data Analysis Software
(NuSTARDAS) jointly developed by the ASI Science Data Center
(ASDC, Italy) and the California Institute of Technology (Caltech, USA)
and the XRT Data Analysis Software (XRTDAS) developed under the
responsibility of ASDC.
T.K. was supported by Japan Society for the Promotion of Science (JSPS)
Grant-in-Aid for Young Scientists (B) (No. 24740185).

{\it Facilities:} \facility{NuSTAR}, \facility{Swift}.

\clearpage

\end{document}